\documentclass[letterpaper,9pt,twocolumn,twoside,]{pinp}

\providecommand{\tightlist}{%
  \setlength{\itemsep}{0pt}\setlength{\parskip}{0pt}}

\usepackage[T1]{fontenc}
\usepackage[utf8]{inputenc}

\definecolor{pinpblue}{HTML}{185FAF}  
\definecolor{pnasbluetext}{RGB}{101,0,0} %

\usepackage{xtab,booktabs}
\usepackage{xifthen}
\DeclareFontFamily{OT1}{pzc}{}
\DeclareFontShape{OT1}{pzc}{m}{it}{<-> s * [1.250] pzcmi7t}{}
\DeclareMathAlphabet{\mathscr}{OT1}{pzc}{m}{it}

\title{What Were They Thinking? Pharmacologic priors implicit in a
choice of 3+3 dose-escalation design}

\author[a]{David C. Norris}

  \affil[a]{Precision Methodologies, LLC, Wayland MA, 01778}

\leadauthor{Norris, DC}

 \keywords{  dose-finding studies |  dose
individualization |  oncology |  trial safety |  precision medicine  }  

\begin{abstract}
If explicit, formal consideration of clinical pharmacology at all
informs the design and conduct of modern oncology dose-finding trials,
the designs themselves hardly attest to this. Yet in conducting a trial,
investigators affirm that they hold reasonable expectations of
participant safety---expectations that necessarily depend on beliefs
about how certain pharmacologic parameters are distributed in the study
population. Thus, these beliefs are implicit in a trial's presumed
conformance to a community standard of safety, and may therefore to some
extent be reverse-engineered from trial designs. For one popular form of
dose-escalation trial design, I demonstrate here how this may be done.
\end{abstract}

\dates{WORKING PAPER DRAFT compiled \today.}

\doifooter{\url{https://cran.r-project.org/package=precautionary}}

\pinpfootercontents{On the priors implicit in a 3+3 design}

\begin{document}

\verticaladjustment{-2pt}

\maketitle
\thispagestyle{firststyle}
\ifthenelse{\boolean{shortarticle}}{\ifthenelse{\boolean{singlecolumn}}{\abscontentformatted}{\abscontent}}{}


\newcommand{\MTD}[2][]{\ensuremath{\ifthenelse{\isempty{#1}}{\mathrm{MTD}_{#2}}{\mathrm{MTD}_{#1}^{#2}}}}
\newcommand{\Figure}[1]{\mbox{Figure \ref{#1}}}
\newcommand{\Table}[1]{\mbox{Table \ref{#1}}}
\newcommand{\plane}[2]{$#1$\nobreakdash--$#2$~plane}

\makeatletter
\def\tagform@#1{\maketag@@@{(\ignorespaces#1\unskip\@@italiccorr)}}
\makeatother

\newcommand{\eqsrange}[2]{Eqs.\ (\ref{#1}--\ref{#2})}

\newcommand{\MTDi}{\mathrm{MTD}_i}
\newcommand{\MTDig}[1][g]{\mathrm{MTD}_i^{#1}}
\newcommand{\CV}{\mathrm{CV}}

\hypertarget{background}{%
\section{Background}\label{background}}

The safety characteristics of a dose-finding study are a function
jointly of how the study is designed, and of how certain pharmacologic
parameters are distributed in the study population. Given an explicit
set of priors over those distributions, one can carry out simulations to
exhibit the safety characteristics of any proposed design: \[
(\mathrm{priors}, \mathrm{design}) \xrightarrow{simulation} \mathrm{safety}
\] We may express this \emph{conditionally} by saying that \emph{given
any trial design,} our safety expectations are a function
\(F_{\mathrm{design}}\) of our priors:
\begin{equation}\label{eq:forward}
\mathrm{priors} \xrightarrow{F_{\mathrm{design}}} \mathrm{safety}.
\end{equation}

In oncology dose finding, however, population heterogeneity is rarely
acknowledged, let alone modeled explicitly through Bayesian
priors.\footnote{Some reasons for this state of affairs may be gleaned
  from \citet{sheiner_intellectual_1991}.} Nevertheless, restrictions on
such priors are \emph{implicit} in whatever bounds we can identify on
acceptable trial safety characteristics. Thus, community standards which
limit the numbers of severe or fatal toxicities acceptable in a given
clinical-trial context provide information about what pharmacologic
priors one could reasonably entertain while proposing the trial design.
In a sense, recovering this information amounts to solving an
\emph{inverse problem}, \begin{align}
\mathrm{priors} &\xleftarrow{F_{\mathrm{design}}^{-1}} \mathrm{safety} \notag
\\
&\quad\mathrm{or} \label{eq:inverse}
\\
\mathrm{priors} &= F_{\mathrm{design}}^{-1} (\mathrm{safety}) \notag
\end{align} corresponding to the `forward problem' of
\eqref{eq:forward}.

\newpage

\hypertarget{sources-of-community-standards}{%
\section{Sources of community
standards}\label{sources-of-community-standards}}

Explicit discussion of standards for early-phase oncology trial safety
are as rare as explicit discussions of pharmacologic priors.\footnote{Indeed,
  when considering ordinal toxicities, the dose-finding literature
  typically excludes fatal toxicities from the universe of possible
  outcomes
  \citep{bekele_dose-finding_2004, van_meter_dose-finding_2012}.} Thus
\eqref{eq:inverse} appears vulnerable to a symmetry argument, to the
effect that it merely presents a mirror-image of the very same
difficulties posed by \eqref{eq:forward}. Both Equations derive one set
of priors from another; they differ only in whether objective
pharmacologic priors or subjective safety priors are taken as the
starting point. Since neither set of priors receives any amount of
explicit discussion, both starting points would seem to be equally
inaccessible.

This criticism is valid inasmuch as it reveals our problem to be one of
\emph{prior elicitation}. But the supposed symmetry between
\eqref{eq:forward} and \eqref{eq:inverse} is broken---on purely
practical grounds---by the primacy of \emph{safety} in drug development.
While it remains (however remarkably) entirely possible to evade
explicit prior elicitation around pharmacologic parameters, it is
politically infeasible to brush aside questions of safety \emph{once
they have been posed}.\footnote{The core aim of this paper is to pose
  this question \emph{effectively}---in objective terms which cannot be
  ignored.}

We see a clear manifestation of this principle in the FDA's reflexive
responsiveness to fatalities in early-phase oncology trials. FDA
guidance on phase 1 dose escalation offers only indefinite suggestions
as to how preclinical pharmacology may shape trial design.\footnote{See,
  for example, Section 3.1 of \citet{fda_cder_good_2013}.} Yet once the
\emph{occurrence} of a fatal toxicity starkly `poses the question', FDA
acts swiftly to place a clinical hold---with provisions existing for
clinical hold orders to be ``made by telephone or other means of rapid
communication'' \citep{fda_cder_ind_2018}.

Standards of safety vary with clinical context---disease severity, unmet
need, competitive environment \citep{muller_determination_2012}.
Accordingly, in this analysis we will treat \emph{safety} as a free
parameter and will focus on elaborating the function
\(F_{\mathrm{design}}^{-1}\). For the sake of definiteness, we will
operationalize \emph{safety} as the (probabilistic) expectation of the
number of fatal toxicities in a trial.\footnote{A metric of this kind
  readily generalizes to expectations of less severe events as well,
  such as \mbox{grade 4} toxicities.}

\hypertarget{seeking-efficiencies}{%
\section{Seeking efficiencies}\label{seeking-efficiencies}}

In order to render the inverse problem \eqref{eq:inverse} feasible, we
first seek efficient means to carry out the forward simulation
\eqref{eq:forward}. To this end, we exploit the enumerablity of the
possible paths that rule-based (`algorithmic') dose-finding designs may
follow,\footnote{I would like to acknowledge
  \citet{sabanes_bove_model-based_2019} for impressing this point upon
  me, specifically in the opening remarks to Section 10 of the 12 June
  2019
  \href{https://cran.r-project.org/web/packages/crmPack/vignettes/example.pdf\#page=47}{introductory
  vignette} to \texttt{crmPack}.} which enables exact computations
exempt from Monte Carlo error. The Prolog program below implements a
definite clause grammar (DCG) that generates all such paths, for the
common variant of the 3 + 3 design in which 6 patients must have been
treated at a dose level before it may be declared `the MTD'
\citep{korn_comparison_1994, skolnik_shortening_2008}.

\newpage

\begin{Shaded}
\begin{Highlighting}[]

\NormalTok{tox(}\DataTypeTok{T}\NormalTok{) }\KeywordTok{:{-}} \DataTypeTok{T}\NormalTok{ in }\DecValTok{0}\NormalTok{..}\DecValTok{3}\KeywordTok{,}
\NormalTok{          indomain(}\DataTypeTok{T}\NormalTok{)}\KeywordTok{.}

\NormalTok{esc(}\DataTypeTok{Hi}\KeywordTok{,}\DataTypeTok{Lo}\NormalTok{..}\DataTypeTok{Hi}\NormalTok{) }\KeywordTok{{-}{-}\textgreater{}}\NormalTok{ [}\DataTypeTok{Hi} \FunctionTok{*} \DataTypeTok{T}\NormalTok{]}\KeywordTok{,} \KeywordTok{\{}\NormalTok{ tox(}\DataTypeTok{T}\NormalTok{) }\KeywordTok{\},}
\NormalTok{                   (  }\KeywordTok{\{}\DataTypeTok{T}\NormalTok{ \#=\textless{} }\DecValTok{1}\KeywordTok{\},}\NormalTok{ [mtd\_notfound(}\DataTypeTok{Hi}\NormalTok{)]}
                   \KeywordTok{;}  \KeywordTok{\{}\DataTypeTok{T}\NormalTok{ \#\textgreater{}= }\DecValTok{2}\KeywordTok{\},}\NormalTok{ des(}\DataTypeTok{Hi}\KeywordTok{,} \DataTypeTok{Lo}\NormalTok{)}
\NormalTok{                   )}\KeywordTok{.}
\NormalTok{esc(}\DataTypeTok{D}\KeywordTok{,} \DataTypeTok{Lo}\NormalTok{..}\DataTypeTok{Hi}\NormalTok{) }\KeywordTok{{-}{-}\textgreater{}} \KeywordTok{\{} \DataTypeTok{D}\NormalTok{ \#\textless{} }\DataTypeTok{Hi}\KeywordTok{,} \DataTypeTok{D1}\NormalTok{ \#= }\DataTypeTok{D} \FunctionTok{+} \DecValTok{1} \KeywordTok{\},}
\NormalTok{                   [}\DataTypeTok{D1} \FunctionTok{\^{}} \DataTypeTok{T}\NormalTok{]}\KeywordTok{,} \KeywordTok{\{}\NormalTok{ tox(}\DataTypeTok{T}\NormalTok{) }\KeywordTok{\},}
\NormalTok{                   (  }\KeywordTok{\{}\DataTypeTok{T}\NormalTok{ \#= }\DecValTok{0}\KeywordTok{\},}\NormalTok{ esc(}\DataTypeTok{D1}\KeywordTok{,} \DataTypeTok{Lo}\NormalTok{..}\DataTypeTok{Hi}\NormalTok{)}
                   \KeywordTok{;}  \KeywordTok{\{}\DataTypeTok{T}\NormalTok{ \#= }\DecValTok{1}\KeywordTok{\},}\NormalTok{ sta(}\DataTypeTok{D1}\KeywordTok{,} \DataTypeTok{Lo}\NormalTok{..}\DataTypeTok{Hi}\NormalTok{)}
                   \KeywordTok{;}  \KeywordTok{\{}\DataTypeTok{T}\NormalTok{ \#\textgreater{} }\DecValTok{1}\KeywordTok{\},}\NormalTok{ des(}\DataTypeTok{D1}\KeywordTok{,} \DataTypeTok{Lo}\NormalTok{)}
\NormalTok{                   )}\KeywordTok{.}

\NormalTok{sta(}\DataTypeTok{D}\KeywordTok{,}  \DataTypeTok{\_}\NormalTok{..}\DataTypeTok{D}\NormalTok{ ) }\KeywordTok{{-}{-}\textgreater{}}\NormalTok{ [}\DataTypeTok{D} \FunctionTok{{-}} \DecValTok{0}\NormalTok{]}\KeywordTok{,}\NormalTok{ [mtd\_notfound(}\DataTypeTok{D}\NormalTok{)]}\KeywordTok{.}
\NormalTok{sta(}\DataTypeTok{D}\KeywordTok{,} \DataTypeTok{Lo}\NormalTok{..}\DataTypeTok{Hi}\NormalTok{) }\KeywordTok{{-}{-}\textgreater{}} \KeywordTok{\{} \DataTypeTok{D}\NormalTok{ \#\textless{} }\DataTypeTok{Hi}\KeywordTok{,} \DataTypeTok{D}\NormalTok{ in }\DataTypeTok{Lo}\NormalTok{..}\DataTypeTok{Hi} \KeywordTok{\},}
\NormalTok{                   [}\DataTypeTok{D} \FunctionTok{{-}} \DecValTok{0}\NormalTok{]}\KeywordTok{,}
\NormalTok{                   esc(}\DataTypeTok{D}\KeywordTok{,} \DataTypeTok{D}\NormalTok{..}\DataTypeTok{Hi}\NormalTok{)}\KeywordTok{.}
\NormalTok{sta(}\DataTypeTok{D}\KeywordTok{,} \DataTypeTok{Lo}\NormalTok{..}\DataTypeTok{\_}\NormalTok{ ) }\KeywordTok{{-}{-}\textgreater{}}\NormalTok{ [}\DataTypeTok{D} \FunctionTok{{-}} \DataTypeTok{T}\NormalTok{]}\KeywordTok{,} \KeywordTok{\{}\NormalTok{ tox(}\DataTypeTok{T}\NormalTok{)}\KeywordTok{,} \DataTypeTok{T}\NormalTok{ \#\textgreater{} }\DecValTok{0} \KeywordTok{\},}
\NormalTok{                   des(}\DataTypeTok{D}\KeywordTok{,} \DataTypeTok{Lo}\NormalTok{)}\KeywordTok{.}

\NormalTok{des(}\DataTypeTok{D}\KeywordTok{,} \DataTypeTok{Lo}\NormalTok{) }\KeywordTok{{-}{-}\textgreater{}} \KeywordTok{\{} \DataTypeTok{D\_1}\NormalTok{ \#= }\DataTypeTok{D} \FunctionTok{{-}} \DecValTok{1} \KeywordTok{\},}
\NormalTok{               (  }\KeywordTok{\{}\DataTypeTok{D\_1}\NormalTok{ \#= }\DataTypeTok{Lo}\KeywordTok{\},}\NormalTok{ [declare\_mtd(}\DataTypeTok{Lo}\NormalTok{)]}
               \KeywordTok{;}  \KeywordTok{\{}\DataTypeTok{D\_1}\NormalTok{ \#\textgreater{} }\DataTypeTok{Lo}\KeywordTok{\},}\NormalTok{ [}\DataTypeTok{D\_1} \FunctionTok{:} \DataTypeTok{T}\NormalTok{]}\KeywordTok{,} \KeywordTok{\{}\NormalTok{tox(}\DataTypeTok{T}\NormalTok{)}\KeywordTok{\},}
\NormalTok{                  (  }\KeywordTok{\{}\DataTypeTok{T}\NormalTok{ \#=\textless{} }\DecValTok{1}\KeywordTok{\},}\NormalTok{ [declare\_mtd(}\DataTypeTok{D\_1}\NormalTok{)]}
                  \KeywordTok{;}  \KeywordTok{\{}\DataTypeTok{T}\NormalTok{ \#\textgreater{}= }\DecValTok{2}\KeywordTok{\},}\NormalTok{ des(}\DataTypeTok{D\_1}\KeywordTok{,} \DataTypeTok{Lo}\NormalTok{)}
\NormalTok{                  )}
\NormalTok{               )}\KeywordTok{.}
\end{Highlighting}
\end{Shaded}

\hypertarget{efficient-simulation-of-33}{%
\subsection{Efficient simulation of
3+3}\label{efficient-simulation-of-33}}

In 3+3 designs, each 3-patient cohort has 1 of 4 possible outcomes,
according to the count of dose-limiting toxicities (DLTs): 0/3, 1/3,
2/3, or 3/3. In the course of a 3+3 trial, each dose may enroll 0, 1 or
2 cohorts. Thus, it is possible to represent the events on path \(j\) by
a \(2 \times D\) matrix \((T_{c,d}^j)\) with rows indexed by cohort
\(c \in \{1, 2\}\) and columns by dose level \(d \in \{1,...,D\}\), and
with elements drawn from \(\{0, 1, 2, 3, -\}\). For example, the matrix
\[
\left(
\begin{array}{c c c c}0 & 1 & 2 & -\\
- & 0 & - & -
\end{array}
\right)
\] represents a path in a 4-dose 3+3 trial, where the following events
occur:

\begin{enumerate}
\def\labelenumi{\arabic{enumi}.}
\tightlist
\item
  Initial cohort at \(d=1\) results 0/3
\item
  Escalation to \(d=2\) results 1/3
\item
  Additional cohort at \(d=2\) results 0/3 for net 1/6 at this dose
\item
  Escalation to \(d=3\) results 2/3; MTD declared at \(d=1\).
\end{enumerate}

The matrices \(T^j\) support concise expression and efficient, exact
computation of trial outcomes and their probabilities. For example, the
\(J\)-vector \((\pi^j)\) of path probabilities may be written\footnote{Products
  or sums over \(c\) or pairs \((c,d)\) are understood to be taken over
  the \emph{non-empty} cohorts which are thus indexed. In R, this
  corresponds to treating the `\(-\)' entries as \texttt{NA} values, and
  employing the convention \texttt{na.rm\ =\ TRUE} in aggregate
  operations.} \begin{align}
p_d &= P(\mathrm{MTD}_i< X_d) \notag \\
q_d &= 1 - p_d \notag \\
\pi^j &= \prod_{c,d} \binom{3}{T_{c,d}^j}\, p_d^{T_{c,d}^j} q_d^{(3-T_{c,d}^j)}, \label{eq:pi}
\end{align} \newpage \noindent where \((p_d)\) is the \(D\)-vector of
DLT probabilities at the prespecified doses \((X_d)\), and
\((q_d) \equiv (1-p_d)\) is its complement. Felicitously, taking logs in
\eqref{eq:pi} yields the matrix equation: \begin{align}
\log \boldsymbol{\pi} &= \sum_{c,d} \log \binom{3}{T_{c,d}} + \sum_c [T_{c,d},3-T_{c,d}] \genfrac{[}{]}{0pt}{}{ \log \mathbf{p} }{ \log \mathbf{q}} \notag \\
&= \mathbf{b}_D + U_D \genfrac{[}{]}{0pt}{}{ \log \mathbf{p} }{ \log \mathbf{q}}. \label{eq:logpi}
\end{align} Observe that the \(J \times 2D\) blocked matrix \(U_D\) and
\(J\)-vector \(\mathbf{b}_D\) thus defined are \emph{characteristic
constants} of the 3+3 design for a given value of \(D\), and that the
distribution \(P\) of \(\mathrm{MTD}_i\) in the population enters
\eqref{eq:logpi} only through the column vector
\([\log \mathbf{p}; \log \mathbf{q}]\).

As shown in \mbox{Table \ref{tab:JvD}}, the number of possible paths
\(J_D\) for this standard 3+3 design grows almost exponentially with the
number of dose levels \(D\). Nevertheless, for trial sizes of practical
interest, say \(D \le 8\), the matrices involved remain trivially small
in terms of computer memory and computation time.

\begin{table}[ht]
\centering
\caption{Number of possible dose-escalation paths in the standard 3+3 design, as a function of the number $D$ of prespecified doses.} 
\label{tab:JvD}
\begin{tabular}{rrrc}
  \hline
$D$ & $J$ & $\log(J)$ & $\Delta \log(J)$ \\ 
  \hline
1 & 10 & 2.30 & -- \\ 
  2 & 46 & 3.83 & 1.53 \\ 
  3 & 154 & 5.04 & 1.21 \\ 
  4 & 442 & 6.09 & 1.05 \\ 
  5 & 1162 & 7.06 & 0.97 \\ 
  6 & 2890 & 7.97 & 0.91 \\ 
  7 & 6922 & 8.84 & 0.87 \\ 
  8 & 16138 & 9.69 & 0.85 \\ 
  9 & 36874 & 10.52 & 0.83 \\ 
  10 & 82954 & 11.33 & 0.81 \\ 
   \hline
\end{tabular}
\end{table}

\hypertarget{ordinalization-of-toxicity}{%
\subsection{Ordinalization of
toxicity}\label{ordinalization-of-toxicity}}

Following \citet{norris_retrospective_2020}, we posit a therapeutic
index \(e^\kappa\) establishing a fixed ratio between dose thresholds
\(\{\mathrm{MTD}_i^g\}_{g=3}^5\) for toxicities of grades
3--5,\footnote{In \citet{norris_retrospective_2020}, this ratio was
  called \(r_0\).} \begin{equation}
\mathrm{MTD}_i^{g}= e^{\kappa{(g-3)}} \cdot \mathrm{MTD}_i, \;\; g \in \{3, 4, 5\}, \label{eq:MTDig}
\end{equation} so that in particular the threshold
\(\mathrm{MTD}_i^{5}\) for fatal toxicity is\footnote{No assumption
  about \(\mathrm{MTD}_i^{4}\) is required strictly for purposes of
  analyzing \emph{fatal} toxicities. The geometric sequence of
  \eqref{eq:MTDig} should therefore be appreciated as a heuristic to
  assist prior elicitation about the quantity \(e^{2\kappa}\) in
  \eqref{eq:fatal-link}, in terms of what may prove for oncologists a
  more intuitive linkage \(e^\kappa = \sqrt{e^{2\kappa}}\) connecting
  \emph{adjacent} toxicity grades.} \begin{equation}
\mathrm{MTD}_i^{5} = e^{2\kappa} \cdot \mathrm{MTD}_i. \label{eq:fatal-link}
\end{equation} Consequently, of all DLTs (toxicities of grade \(\ge\) 3)
occurring at dose \(X_d\), the fatal (\mbox{grade 5}) fraction \(f_d\)
is: \[
f_d = \frac{P(e^{2\kappa} \mathrm{MTD}_i<  X_d)}{P(\mathrm{MTD}_i< X_d)}.  \label{eq:fatalfraction}
\] The left half of \(U_D\) is a \((J \times D)\) matrix
\(Y_D = (y_d^j) = (\Sigma_c T_{c,d}^j)\) which is itself of interest,
since its \(j\)th row tells how many DLTs occur at each dose on the
\(j\)th path of the trial. In terms of \(Y\), we may write the expected
number of fatal toxicities as:

\[
\boldsymbol{\pi}^\intercal \mathrm{Y} \mathbf{f}.  \label{eq:fatalities}
\]

\begin{figure*}
  \begin{center}
    \includegraphics[width=\textwidth, keepaspectratio]{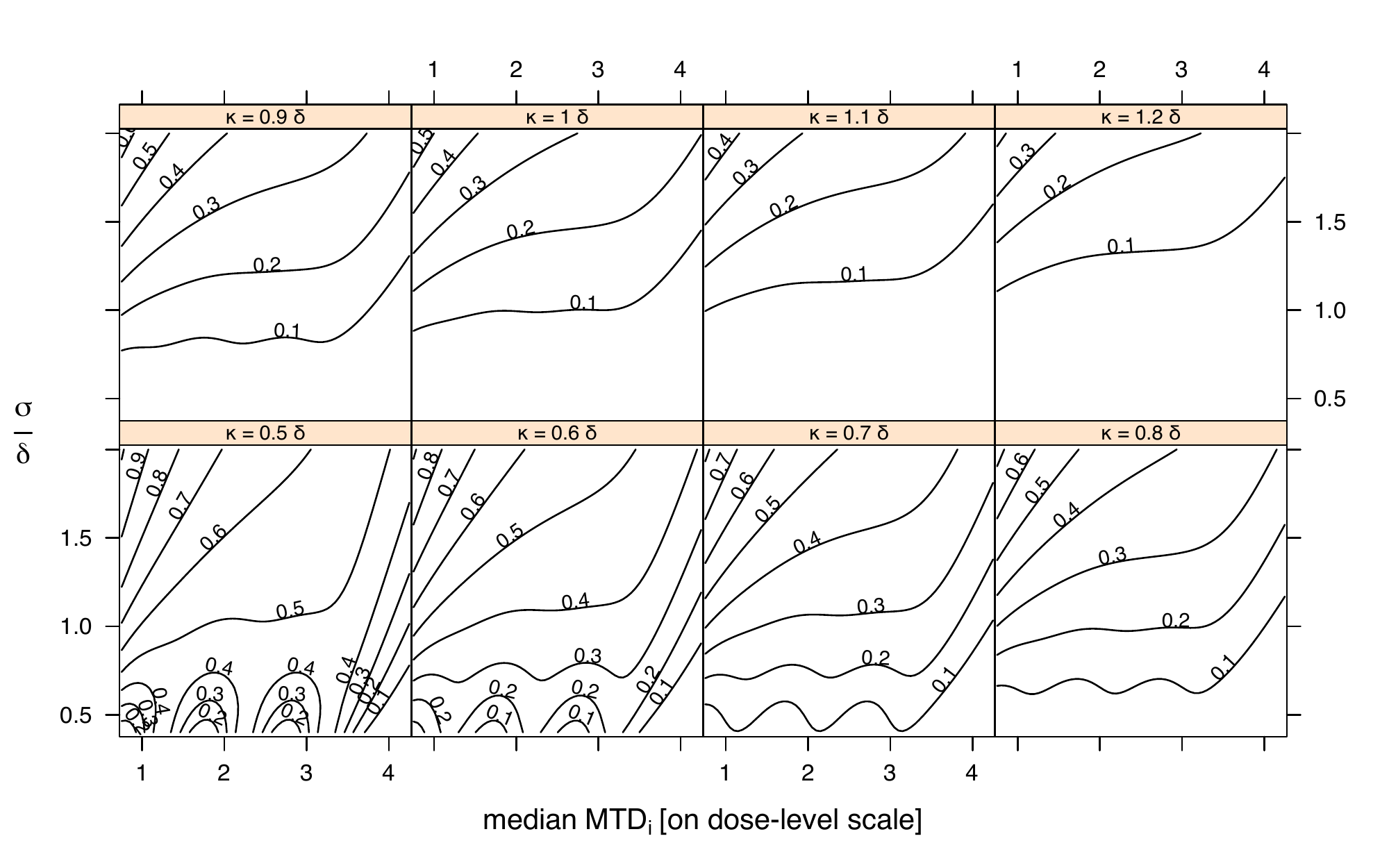}
    \includegraphics[width=\textwidth, keepaspectratio]{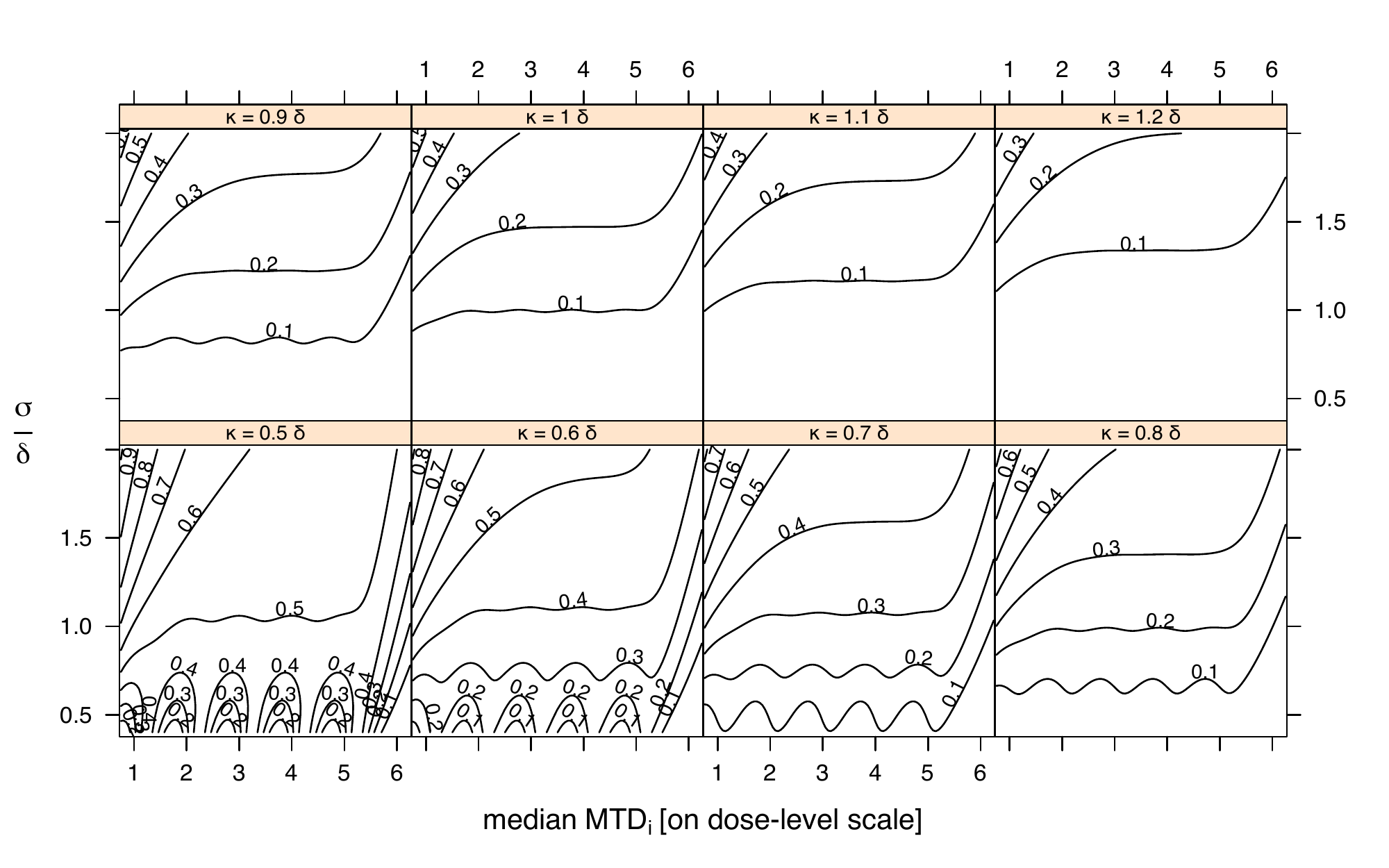}
  \end{center}
  \caption{Expected fatal toxicities for a 3+3 dose-escalation trial with 4 (top) or 6 (bottom) pre-specified doses, under a range of scenarios for log-therapeutic index $\kappa$ and for the parameters $(\mu, \sigma)$ governing the log-normal distribution of $\mathrm{MTD}_i$ in the population: $\log \mathrm{MTD}_i\sim \mathscr{N}(\mu, \sigma)$. The pre-specified doses are assumed to be in a geometric sequence with ratio $e^\delta$. The therapeutic index $e^\kappa$ is the dose multiplier that worsens an individual's experienced toxicity by 1 grade level.}\label{fig:contours46}
\end{figure*}

\newpage

\hypertarget{solving-the-inverse-problem}{%
\section{Solving the inverse
problem}\label{solving-the-inverse-problem}}

We will take a graphical approach to solving the inverse problem
\eqref{eq:inverse}, plotting a graph of \(F_{\mathrm{design}}\) in such
a way that crucial aspects of its inverse become visually accessible.

\hypertarget{treating-the-dose-domain-logarithmically}{%
\subsection{Treating the dose domain
logarithmically}\label{treating-the-dose-domain-logarithmically}}

A resolutely logarithmic treatment of the dose space enables a
substantial reduction in dimensionality for our problem. Accordingly, we
suppose that \(\mathrm{MTD}_i\) is log-normally distributed: \[
\log \mathrm{MTD}_i\sim \mathscr{N}(\mu, \sigma),
\] and we require that our design's prespecified doses \((X_d)\) be
spaced logarithmically at fixed intervals of \(\delta\): \[
X_d = e^\delta X_{d-1}; \quad \log X_d = \delta + \log X_{d-1}.
\] The safety function \(F_{\mathrm{design}}\) may be regarded as a
function of pharmacologic parameters \((\mu, \sigma, \kappa)\): \[
F_{D,\delta}(\mu, \sigma, \kappa),
\] parametrized by the design \((D, \delta)\).

\hypertarget{natural-scales-for-mathbfmu-mathbfsigma-and-mathbfkappa}{%
\subsection{\texorpdfstring{Natural scales for \(\mathbf{\mu}\),
\(\mathbf{\sigma}\) and
\(\mathbf{\kappa}\)}{Natural scales for \textbackslash mathbf\{\textbackslash mu\}, \textbackslash mathbf\{\textbackslash sigma\} and \textbackslash mathbf\{\textbackslash kappa\}}}\label{natural-scales-for-mathbfmu-mathbfsigma-and-mathbfkappa}}

Without loss of generality, \(\mu\) can be measured against the
logarithmic scale generated by our dose indexes:
\(\mu' = 1, 2, ..., D\). On this understanding, \(F\) turns out to be
\emph{invariant} to transformations of \((\delta, \sigma, \kappa)\) that
preserve the ratios \(\sigma' = \sigma/\delta\) and
\(\kappa' = \kappa/\delta\). This enables the design parameter
\(\delta\) to be factored out as a natural scale for measuring
\(\sigma\) and \(\kappa\): \[
F_{D,\delta}(\mu,\sigma,\kappa) = F_D(\mu',\sigma',\kappa').
\] The safety function \(F_D\) then becomes a scalar field in
\(\mathbb{R}^3\), amenable to a treatment as in
\mbox{Figure \ref{fig:contours46}}, where contours in the
$\mu'$\nobreakdash--$\sigma'$~plane are plotted for
\(\kappa' \in \{0.5, 0.6,...,1.2\}\), and \(D \in \{4,6\}\).

Already \mbox{Figure \ref{fig:contours46}} begins to reveal certain
rectangular constraints sufficient to ensure a reasonable standard of
safety. For example, \(\sigma < \delta < \kappa\) apparently keeps the
probability of a fatal toxicity below 0.1 in 3+3 trials with 4 to 6 dose
levels, so long as the dose range includes the median
\(\mathrm{MTD}_i\). Still, amidst the 3 dimensions of this figure, a
comprehensive delineation of safe regions in the design-parameter space
remains elusive.

\hypertarget{further-dimension-reduction-via-minimax}{%
\subsection{Further dimension reduction via
minimax}\label{further-dimension-reduction-via-minimax}}

\begin{figure}

{\centering \includegraphics{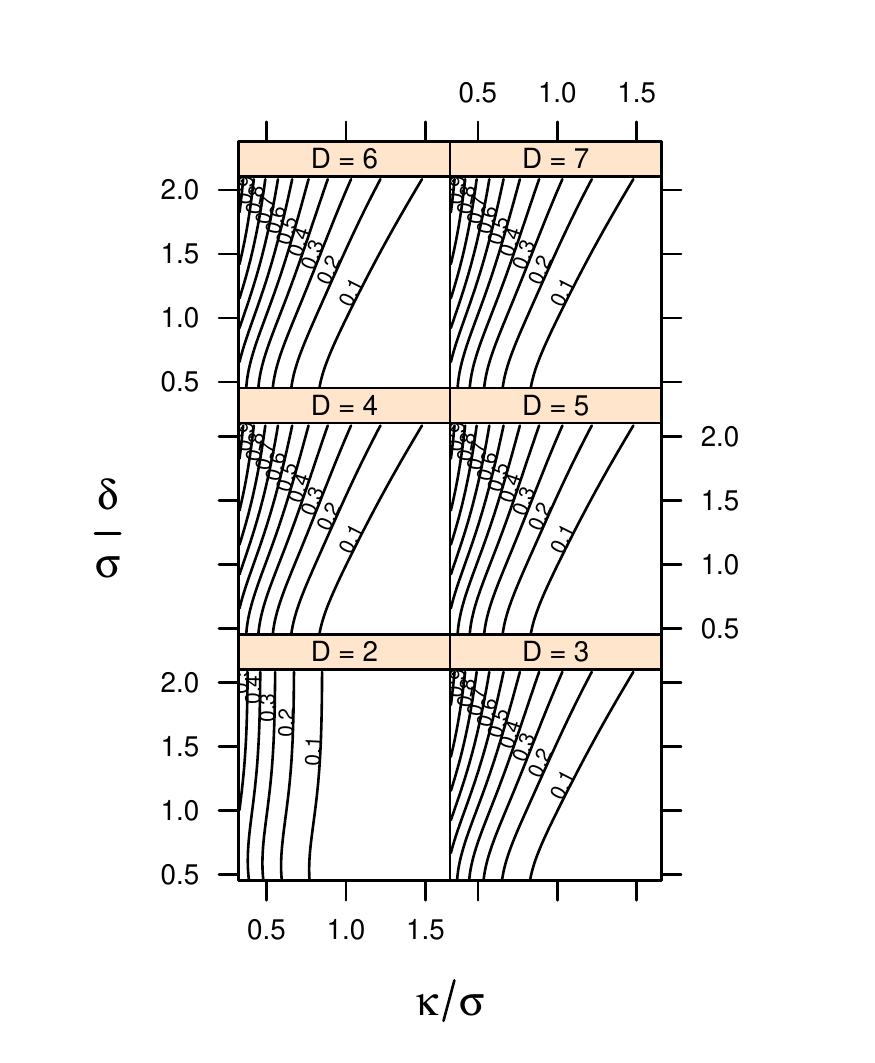} 

}

\caption{Contours of expected number of fatal toxicities in 3+3 trials with 2--7 prespecified doses, assuming---as a `worst-case scenario'---that median $\mathrm{MTD}_i$ equals dose level 2. Remarkably, for $D \ge 3$ this intuitive minimax scenario construction (taken together with the other scalings employed here) brings these figures into almost perfect coincidence.}\label{fig:focused-inversion}
\end{figure}

By offering a \emph{minimax} framing for our safety considerations,
however, we may eliminate yet another dimension. Key to achieving this
framing plausibly, is recognizing that a rational argument in favor of
any given design must be hierarchically structured. At the base of the
hierarchy will be the \emph{sine qua non} that our starting dose is
safe, which in general requires that it lie comfortably below the median
\(\mathrm{MTD}_i\). But if we allow, as a worst-case scenario, that
median \(\mathrm{MTD}_i\) may sit as low as dose level 2, then we can
focus attention on vertical \(\mu' = 2\) `slices' of the panels in
\mbox{Figure \ref{fig:contours46}}: \[
F_D(\mu' = 2,\sigma',\kappa') = F_{D,\,\mu' = 2}(\sigma',\kappa'),
\] which for fixed \(D\) leaves just 2 dimensions. Moreover---and quite
remarkably---this particular choice \(\mu' = 2\) happens to render \(F\)
independent of \(D\) for \(D \ge 3\), as shown in
\mbox{Figure \ref{fig:focused-inversion}}.

\begin{figure}

{\centering \includegraphics{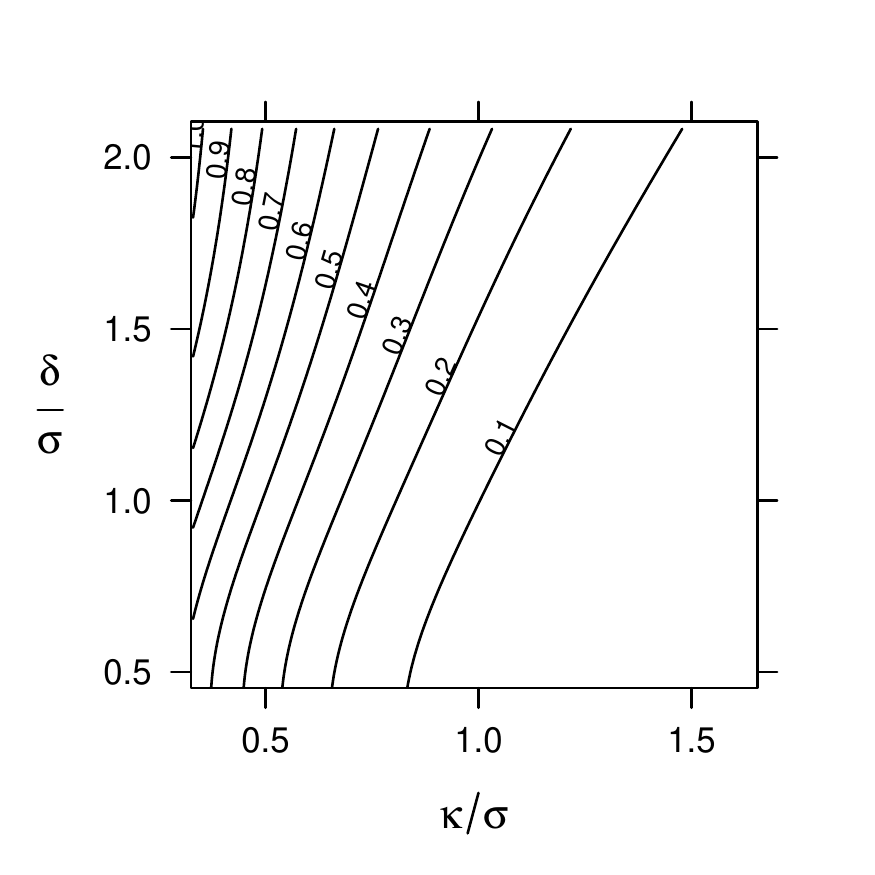} 

}

\caption{Universal design constraints on 3+3 trial safety. Contours for the expected number of fatal toxicities in a 3+3 dose-escalation trial are plotted against 2 crucial design indices. Therapeutic index $\kappa/\sigma$ gauges a drug's aptness for safe-and-effective 1-size-fits-all dosing. Signal-to-noise index $\delta/\sigma$ governs the informativeness of the dose-escalation process.}\label{fig:generic-design}
\end{figure}

Consequently, it is possible to offer a generic safety schematic for 3+3
designs irrespective of the number of prespecified doses, as in
\mbox{Figure \ref{fig:generic-design}}.

\clearpage

\hypertarget{interpreting-the-generic-safety-schematic}{%
\section{Interpreting the generic safety
schematic}\label{interpreting-the-generic-safety-schematic}}

The quantities plotted on the axes of
\mbox{Figure \ref{fig:generic-design}} have intuitive meanings that
greatly facilitate this plot's interpretation and use. The fundamentally
different characters of these quantities moreover suggest a natural
sequence in which they should be considered. Whereas the horizontal axis
is a strictly pharmacological parameter relating two characteristics
\((\kappa, \sigma)\) belonging to \emph{the drug itself,} the vertical
axis incorporates a feature \(\delta\) of the trial design. Since design
logically follows pharmacology, let us consider \(\kappa/\sigma\) first.

\hypertarget{mathbfkappasigma-as-suitability-for-1-size-fits-all-dosing}{%
\subsection{\texorpdfstring{\(\mathbf{\kappa/\sigma}\) as suitability
for 1-size-fits-all
dosing}{\textbackslash mathbf\{\textbackslash kappa/\textbackslash sigma\} as suitability for 1-size-fits-all dosing}}\label{mathbfkappasigma-as-suitability-for-1-size-fits-all-dosing}}

Consider that \(\sigma\) is the population standard deviation of
\(\log \mathrm{MTD}_i\), whereas \(\log \mathrm{MTD}_i+ 2\kappa\) is
individual \(i\)'s fatal dose threshold. For values of
\(\kappa/\sigma \approx 0.5\) we have \(\sigma \approx 2\kappa\), which
places \emph{population-level} variation in optimal dosing on par with
the \emph{individual-level} safety margin separating optimal from fatal
dosing. Clearly, this undermines the feasibility of safe-and-effective
1-size-fits-all dosing. Only when \(\sigma \ll 2\kappa\) may we hope to
find a 1-size-fits-all dose that brings most individuals in the
population within reach of their \(\mathrm{MTD}_i\)'s while
simultaneously ensuring fatal overdoses remain rare. Thus
\(\kappa/\sigma\) gauges a drug's suitability for 1-size-fits-all
dosing. The horizontal axis of \mbox{Figure \ref{fig:generic-design}}
therefore points us to one of the very first considerations we must make
in the clinical development of a drug.

\hypertarget{mathbfdeltasigma-measures-a-signal-to-noise-balance-in-dose-escalation}{%
\subsection{\texorpdfstring{\(\mathbf{\delta/\sigma}\) measures a
signal-to-noise balance in dose
escalation}{\textbackslash mathbf\{\textbackslash delta/\textbackslash sigma\} measures a signal-to-noise balance in dose escalation}}\label{mathbfdeltasigma-measures-a-signal-to-noise-balance-in-dose-escalation}}

The vertical axis in \mbox{Figure \ref{fig:generic-design}} measures the
design's dose increments in units of \(\sigma\). When
\(\delta/\sigma \ll 1\), the effect of a \(1\delta\) dose escalation on
observed toxicities will be swamped by the \(\pm \sigma\) noise
introduced by random enrollment from the population. Thus we may see
\(\delta/\sigma\) as a signal-to-noise index for dose escalation.
\mbox{Figure \ref{fig:generic-design}} makes clear that to increase
signal-to-noise in a dose-escalation trial, we must pay a price in
safety---a price that may be exorbitant for \(\kappa/\sigma < 1\).

\hypertarget{critical-application-of-the-safety-schematic}{%
\section{Critical application of the safety
schematic}\label{critical-application-of-the-safety-schematic}}

Notwithstanding the logical dictum that pharmacology precedes design,
\mbox{Figure \ref{fig:generic-design}} will more likely see immediate
and impactful applications in the reverse direction---as a device for
\emph{criticizing} proposed designs in the manner of \eqref{eq:inverse}.
Starting from the \(\delta\) of a given 3+3 design, one may ask a trial
sponsor what signal-to-noise index \(\delta/\sigma\) is being targeted,
and then how the implied value of \(\sigma\) compares with the safety
margin \(\kappa\) as estimated from preclinical studies or from previous
clinical experience with the drug class. Finally, noting the expected
number of fatal toxicities indicated by the schematic, the sponsor
should justify these as appropriate to the therapeutic context. Thus,
\mbox{Figure \ref{fig:generic-design}} could lend definite focus to
regulators' and IRBs' critical scrutiny of dose-escalation designs.

\hypertarget{extensions-of-the-approach}{%
\section{Extensions of the approach}\label{extensions-of-the-approach}}

The techniques used here will apply to all dose-escalation designs whose
enumerable outcomes render exact computations like \eqref{eq:logpi}
feasible. Developments such as \emph{dose transition pathways}
\citep{yap_dose_2017} apparently extend the applicability of this
approach even to model-based designs. Perhaps no dose-escalation method
driven by binary toxicities can evade this form of analysis.

Further work should seek to characterize dose-finding \emph{accuracy} as
a function of the signal-to-noise index \(\delta/\sigma\). Inaccurate
dose selection has pharmacoeconomic consequences
\citep{norris_costing_2017, norris_one-size-fits-all_2018} which no
fully rational account of dose-escalation trial design can overlook. A
fully worked-out theory of \mbox{Figure \ref{fig:generic-design}} may
well reveal that dose-escalation methods place the safety of trial
participants and the economics of drug development into essential
conflict.

\hypertarget{data-availability}{%
\subsection{Data availability}\label{data-availability}}

Code for reproducing all of this paper's Figures and analyses may be
found at \href{https://osf.io/ye42p/}{doi:10.17605/osf.io/ye42p}.

\hypertarget{competing-interests}{%
\subsection{Competing interests}\label{competing-interests}}

The author operates a scientific and statistical consultancy focused on
precision-medicine methodologies which are surreptitiously advanced by
this article.

\hypertarget{grant-information}{%
\subsection{Grant information}\label{grant-information}}

No grants supported this work.


\bibliography{precautionary-symlink,packages}
\bibliographystyle{plainnat}

\end{document}